\def\ii{\'{\i}}

\def\beg{\begin{equation}}
\def\fim{\end{equation}}
\def\delh{\frac{\partial h(x,t)}{\partial t}}

\documentclass{elsart}

\usepackage{epsfig}

\usepackage{amssymb}

\begin{document}

\begin{frontmatter}

\title{Lattice Model for Approximate Self-Affine Soil Profiles}

\author[label1]{A.P.F. Atman},
\ead{atman@fisica.ufmg.br} 
\ead[url]{www.fisica.ufmg.br/~atman}
\author[label2]{J.G. Vivas Miranda},
\ead{vivas@mail2.udc.es}
\author[label2]{A. Paz Gonzalez},
\author[label1]{J.G. Moreira},

\address[label1]{Departamento de F\ii sica, Instituto de Ci\^encias Exatas,
Universidade Federal de Minas Gerais, C. P. 702
30123-970, Belo Horizonte, MG - Brazil}
\address[label2]{Instituto Universitario de Xeolox\ii a ``Isidro Parga Pondal'',
Universidade da Coru\~na, La Coru\~na, Spain}

\begin{abstract}
A modeling of the soil structure and surface roughness by means of the concepts 
of the fractal growth is presented. Two parameters are used to control
the model: the fragmentation dimension, $D_f$, and the maximum mass
of the deposited aggregates, $M_{max}$. The fragmentation dimension is related to
the particle size distribution through the relation $N(r \ge R) \sim R^{D_f}$,
where $N(r \ge R)$ is the accumulative number of particles with radius greater
than $R$. The size of the deposited aggregates are chose following the power 
law above, and the morphology of the aggregate is random selected using a bond
percolation algorithm. The deposition rules are the same used in the model of 
solid-on-solid deposition with surface relaxation. A comparison of the model 
with real data shows that the Hurst exponent, $H$, measured {\it via} 
semivariogram method and detrended fluctuation analysis, agrees in statistical 
sense with the simulated profiles.
\vspace{1pc}
\end{abstract}

\begin{keyword}
surface roughness \sep soil modelling \sep  self-affine profiles
\PACS 05.45.Df \sep 83.70.Fn\sep 92.40.Lg
\end{keyword}
\end{frontmatter}

\section{Introduction}
The concepts of the fractal geometry have been widely used to describe and 
quantify irregularity in nature, and statistical self-affine properties 
have recently been identified in various Earth's terrain \cite{dietler} and 
profiles \cite{matsu}. Several works in soil science incorporates the 
fractal geometry to des\-cri\-be and modeling soil physical properties, soil 
physical processes and quantify the soil spatial variability 
\cite{yordanov,martin,perfect,gimenez}.

A very important property closely related to the fractal geometry is the soil 
surface roughness, defined as the configuration of the soil microrelief. 
The soil surface roughness exerts great influence on water infiltration, 
erosion and run-off effects. Its quantification is important for understanding 
the soil behavior during degradation processes like rainfall erosion or abrupt 
changes such as those induced by tillage \cite{garcia}. In the last years, a 
considerable effort was done to simulate 
the soil structure; several models propose the si\-mu\-la\-tion of the particle size 
distributions \cite{martin}, the soil surface roughness 
\cite{yordanov,chadoeuf}, the morphology of the pore-solid structure
\cite{perrier}, etc. The major pro\-per\-ties considered in these models are the 
fractal dimension of the soil surface, $D$, the particle size distribution 
(PSD), the pore size distribution and the surface roughness. 

In this work we present a simple lattice model to simulate the soil structure
and reproduce the surface soil roughness. We basically use the ideas of 
the fractal growth models \cite{barab} to generate an approximate self-affine 
profile, that exhibit, as much as possible, a similitude with the fractal 
properties of real soils. It is approximate in the sense that the scaling 
properties of these profiles are valid only in a limited range of scales.
Therefore, the model was validated by means of a comparison between the fractal 
dimension estimated for simulated profiles and for natural soil surfaces. 
The major improvements in our model are the power law distribution of the 
sizes of 
the deposited particles or aggregates and a random selection in the 
allowed morphologies of the aggregates. This two features, not improved 
before by any model, are responsible for the soil structure in our model. 
In the section II we present a detailed description of the model, showing the 
particle size distribution and explaining the algorithm for the random choice 
of the morphologies. We also present a brief summary of the theory of growth 
surfaces. In the section III is presented the results for several maximum 
aggregates sizes and different fragmentation dimensions.
Finally, in the section IV we present our conclusions and perspectives.

\section{The Model}
The motivation in elaborate this model lies in aptitude to simulate some of the
majors properties of agricultural soil, like self-affine profiles and porous
medium. The model have to allow the possibility to change some parameters, as 
the aggregate size distribution, and the maximum mass of the aggregates. These 
two parameters try to cover the structural variability found in natural soils. 
The aggregate size distribution is assumed to have power law behavior 
\cite{perfect,gimenez}, and their exponent, $D_f$, is one of the parameters of 
the model. The other parameter, $M_{max}$, is the maximum mass of particles 
utilized to generate the aggregate, which is approximately related with the 
square of maximum aggregate radius. In spite of others classical lattice models, 
where the particles generated always have the same morphology, this model 
simulate the variability observed in natural soil, generating aggregates by 
a bond percolation algorithm. All this characteristics have to be integrated 
in a very simple model to allow the simulation of relatively large systems 
($L=1000$).

In this way, the major improvement to choose a lattice model is the simplicity 
and the velocity to run the code, what possibilities the simulation of large 
systems with great quantity of deposited particles (hundred millions).
This massive number of deposited particles are needed in order to avoid the 
effect of the constant attachment of micro random variability, represented by 
the random morphology of the aggregates. Another advantage to work in the 
lattice, is the theoretical background furnished by the fractal growth theory, 
presented by Barab\'asi and Stanley \cite{barab}.

\begin{figure} 
\centering
\epsfxsize=7.0cm {\epsfbox{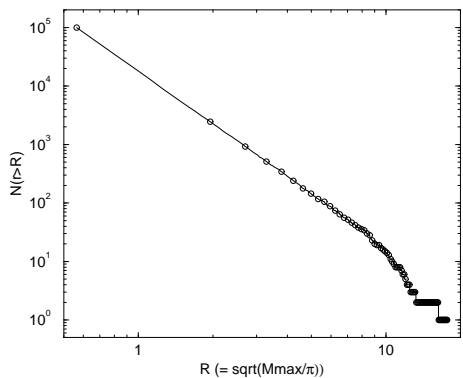}}
\caption{Particle size distribution for a simulated soil with $M_{max} = 1000$
and $D_f = 3.0$.
The radius of the particles are calculeted {\it via} the expression: 
$r^2={M/\pi}$, that approximates the particles by discs with area $M$. Note 
the power law behavior for the particles sizes.}
\end{figure}

We consider the soil structure composed by a set of particles and aggregates 
whose radius are power law distributed according to the following relation, 
known as Turcotte's empirical law \cite{perfect},
\beg\label{turcotte}
N(r > R) \sim R^{-D_f}~~,
\fim
where $N(r>R)$ is the cumulative number of particles (or aggregates) with 
radius $r$ greater than $R$ and $D_f$ is the fragmentation dimension of the 
particle size distribution (PSD). There is some controversy in the literature 
about the allowed range of the values of the $D_f$. Tyler \& Wheatcraft 
\cite{tyler} argued that $D_f < 3$ because, under usual hypothesis (constant 
density, spherical particles, etc), the mass distribution scales with
$M(r>R) \sim R^{D} \sim R^{3-D_f}$,
where $D$ is the fractal dimension of the soil. Thus, considering a fractal 
distribution, only the values $D_f < 3$ have physical meaning. 
Martin \& Taguas \cite{martin} present several mathematical arguments supporting
this conjecture, and show some PSD simulations. However, there are 
several experimental studies, summarized by Perfect \& Kay \cite{perfect}
where the range of values of $D_f$ is $2.6 < D_f < 3.5$. Gimenez {\it et al} 
\cite{almaras} also affirm that there is not any experimental relation between 
the fractal dimension and the fragmentation dimension. We consider that the 
usual hypothesis of constant density of the particles is not valid when the 
radius of the aggregates grows up, due the presence of pores in the structure. 
Thus, fragmentation dimensions greater than 3 are, in principle, allowed. 

According to the USDA classification of soil texture \cite{nemes}, the basic 
particle size classification is: \\
$\bullet$ sand  $50 < r < 2000  \mu m$ \\
$\bullet$ silt  $2 < r < 50  \mu m$ \\
$\bullet$ clay  $r < 2  \mu m$. \\
The real soils can vary widely in the percentile of each range of sizes.
The study of Nemes {\it et al} proposes a standardisation of the classification 
of the European soils and shows several experimental data \cite{nemes}. In the
figure 1, we present a typical PSD generated by the algorithm of the model. This
PSD is constructed considering that the particles are approximately circles; so, 
there is a direct relationship between the mass number and the particle 
radius, that is used to build the PSD.

The model has two parameters, the fragmentation dimension, $D_f$, and the
maximum mass of the aggregates, $M_{max}$. To every particle is selected
a mass number, $M$; when $M>1$, the model choose a random configuration for it; 
this choice is one of the possible bond percolation clusters with a given size. 
In the figure 2, we show some of the possible aggregate morphologies. These two 
parameters try to cover, in a simple way, the structural variability found in 
natural soils.

The deposit of each particle follows the model of the solid-on-solid deposition
with surface relaxation \cite{barab}. The initial position for the deposition 
is random chose and the particle follows a straight line until touch the soil 
surface. Then, the algorithm simulate a surface relaxation (without rotation)
to the particles, which only are adhered to the bulk when they reach the site 
with the local minimum energy .

Considering the theory of growth surfaces, the solid-on-solid deposition
belongs to the universality class of the Edwards-Wilkinson equation
\beg\label{ew}
\delh = \nu \nabla^{2}h + \eta(\vec{x},t)~~~,
\fim
where $h(\vec{x}, t)$ is the height of the site $\vec{x}$ at time $t$, and
$\eta (\vec{x},t)$ represents a random noise, that expresses the fluctuations
in the arrive of the particles and had the following properties,
$<\eta(\vec{x},t)>=0$ and $<\eta(\vec{x},t)~\eta(\vec{x'},t')> 
~=~ 2C~\delta^{d}(\vec{x} - \vec{x'})~\delta(t-t')$.The $\delta$ is 
the usual Kronecker delta and $C$ is a diffusion constant.

\begin{figure} 
\centering
\epsfxsize=8.0cm {\epsfbox{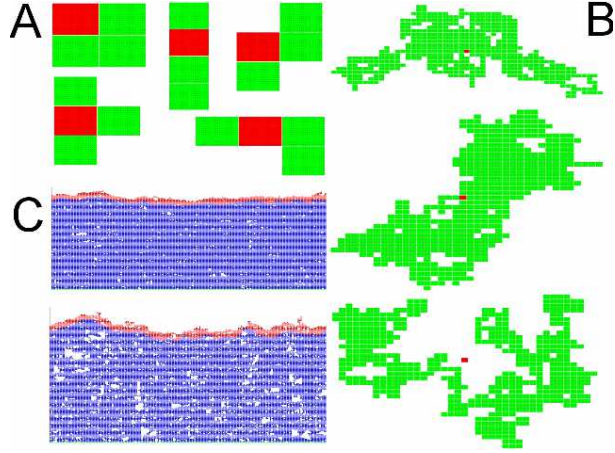}}
\caption{A) Possible morphologies for $M = 4$. In this case, the morphologies
are the same observed in a popular game called ``Tetris". B) Some morphologies
for $M=400$. Three morphologies are shown. Note the porous structure of the 
aggregates, and its random shape. C) Soils profiles generated by the model. 
Above: typical deposition in a lattice with $L=200$, $20000$ particles deposited
with $M_{max}=100$ and $D_f = 3.0$. Below: the same, with $D_f = 2.0$. Note 
the variability of the soil structure with the fragmentation dimension.}
\end{figure}

The width of the growth surfaces, $w(L, t)$ then obeys a scaling relation
\beg\label{scal}
w(L, t) \sim L^{\alpha} f(t/L^z)~~,
\fim
where $f(u)$ is a scaling function: $f(u) \sim u^{\beta}$ for $u<<1$ and
$f(u)$ is a constant for $u>>1$; $\alpha$, $\beta$ and $z$ are the roughness, 
the growth and the dynamical exponents, respectively. They are known as the 
scaling exponents and are related by the scaling law 
\[
z=\frac{\alpha}{\beta}~~.
\] 
The EW scaling exponents are known exactly: $\alpha = 1/2$, $z = 2$.

Another equation related to the deposition is the Kardar-Parisi-Zhang equation, 
that has a nonlinear term,
\beg\label{kpz}
\delh = \nu \nabla^{2}h + \lambda/2 (\nabla h)^2 + \eta(\vec{x},t) ~~~.
\fim
To this equation the calculeted exponents are $\alpha = 1/2$, $z = 3/2$.

So, we perform simulations to test the influence of the model parameters 
$D_f$ and $M_{max}$ in the scaling exponents and fractal dimension of the soil 
profiles, and compare the results with the experimental data disposable.

\begin{figure} 
\centering
\epsfxsize=8.0cm {\epsfbox{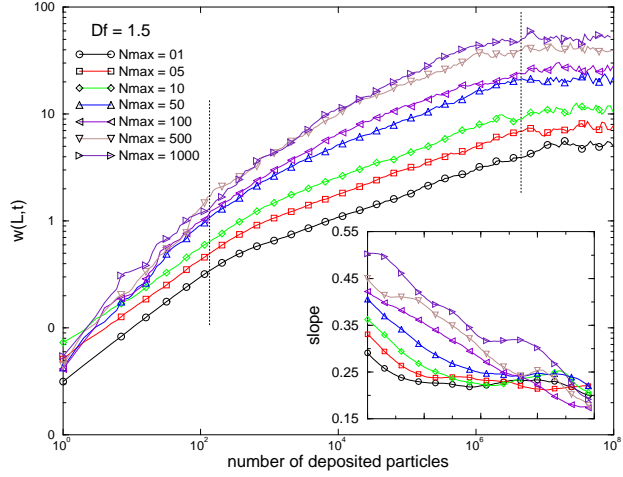}}
\caption{Roughness of the simulated profiles, with several $M_{max}$ and
$D_f = 3.0$. Note the variation of the slope between the two vertical dotted 
lines, shown in the small window. The slope is measured with a rule with fixed
length, and varies the initial point since the left vertical line to the right.}
\end{figure}

\section{Results}
We perform simulations in one dimensional lattice with length $L=1000$ sites.
The results for the scaling exponents represent the average value in a set of 
20 samples with the same parameters in each simulation, and different 
deposition sequence of the random aggregates.
The number of particles/aggregates deposited in each sample is $N=100 000 000$.
The behavior of the local width, $w(L,T)$, defined by
\beg\label{rugos}
w^{2}(L,t) = \frac{1}{L} \sum^{L}_{i=1} \left( h_{i}(t) - \overline{h}(t) 
\right)^{2}~~,
\fim
with the model parameters are shown in the fi\-gu\-re 3. Note the roughness 
saturation at  
$N \sim 10 000 000$. The fractal dimension of the final profile is estimated 
using the Hurst exponent $H$. The Hurst exponent is associated to the fractal 
dimension {\it via} the relation $D = 2 - H$. There are several ways to measure
$H$. At this work, we use the detrended fluctuation analysis (DFA), improved 
by the first time by Moreira {\it et al} \cite{moreira}, and the semivariogram 
method, that uses a height-height correlation function. 

The DFA consists in measure the roughness around the mean square straight 
line \cite{moreira}. The roughness $W(L,\epsilon,t)$ at the scale $\epsilon$, 
is given by
\beg
W(L,\epsilon,t) = \frac{1}{L} \sum_{i=1}^{L} w_i(\epsilon,t)
\fim
\noindent and the local roughness $w_i(\epsilon,t)$ is defined by
\beg
w_i^2(\epsilon,t)=\frac{1}{2\epsilon +1} \sum_{j=i-\epsilon}^{j=i+\epsilon}
\{h_j(t) - [a_i(\epsilon)x_j + b_j(\epsilon)]\}^2
\fim
\noindent where $a_i(\epsilon)$ and $b_i(\epsilon)$ are the linear fitting 
coefficients to the displacement data on the interval $[i-\epsilon, i+\epsilon]$
centered at the site $i$. Self affine profiles satisfy the scaling law
\beg
W(\epsilon) \sim \epsilon^H
\fim
\noindent that is used to measure $H$.
The semivariogram method estimates the spatial variability, through the 
calculus of the semivariance as a function of the distance between points. 
This function is can be estimated by:
\beg\label{semiv}
\gamma(\epsilon, t) = \frac{1}{2 n(\epsilon)} \sum_{i=1}^{n(\epsilon)} 
[h_i(t) - h_{i+\epsilon}(t)]^2 ~~,
\fim
where $h_i(t)$ the height in the location $i$ at time $t$, and $n(\epsilon)$ 
represents the number of pairs of points which are separated by $\epsilon$. 
In the case of self-affine profiles $\gamma$ exhibit a power law behavior,
\[
\gamma(\epsilon) \sim \epsilon^{2H}~~,
\]
and, in the same manner as in the DFA method, the $H$ exponent can be 
utilized to calculate the fractal dimension.

The semivariogram method is especially useful in cases where the data are not
regular spaced, and for this reason, is widely used in estimating fractal 
dimension of soil surfaces \cite{perfect}.

\begin{figure} 
\centering
\epsfxsize=7.0cm {\epsfbox{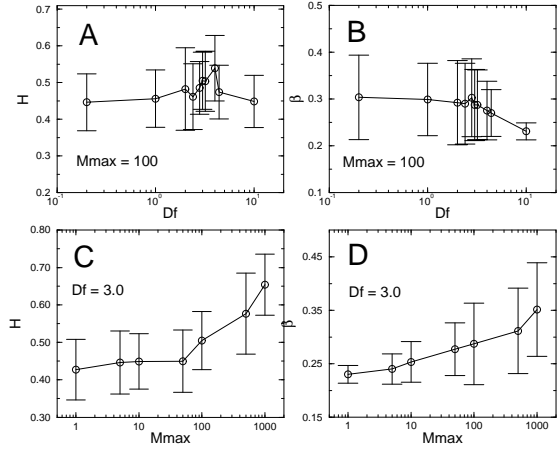}}
\caption{Summary of the exponent values. A) and C): Behavior of the Hurst 
exponent with the model parameters. Note that there is not a remarkable trend
to the Hurst exponent with any model parameters. The range of experimental 
values is $0.3 < H < 0.7$. B) and D): Behavior of the growth exponent
with the model parameters. Note the dependence of the growth exponent with the 
average radius size: as the radius increase, the growth exponent increases too. 
The dependence of $\beta$ with the fragmentation dimension is inverse. }
\end{figure}

The range of the $H$ values is shown in the figure 4A and 4B. In the figure 4C
and 4D, we show 
the dependence of the growth exponent with the model parameters. We conclude 
that the increase of the maximum particle mass $M_{max}$ have similar effects 
to the decrease of the fragmentation dimension: both introduces a nonlinear 
correlation in the system, expressed by the increase in the value of the 
growth exponent $\beta$, changing the universality class of the deposition. 
Nevertheless, the fractal dimension of the surface do not alters considerably
with the model parameters.

\section{Conclusions and Perspectives}

In this work, we present a new model to simulate the soil structure and 
its surface roughness. Two parameters are used to control the simulated 
profiles: the maximum mass of the particles and the fragmentation dimension. 
The major improvements of this model are: the random configuration allowed to
the particles or aggregates and the power law distribution of its radius. These
two features are not present in any model discussed in the literature until 
today, and permits the reproduction of the variability observed in natural 
soils. The results obtained shows a good agreement for the exponent $H$ 
calculated from the simulations and measured experimentally \cite{garcia}. 
We also observe a
dependence of the growth exponent $\beta$ with the maximum mass allowed to the 
particles: as the $M_{max}$ grows, the value of the $\beta$ exponent grows from
the EW value ($\beta = 1/4$) to the KPZ value ($\beta = 1/3$). So, the 
increase of the averaged particle radius corresponds to the introduction of 
nonlinear correlations into the system.

Next, we expect increase the system size and control better the shape of the 
deposited particles, in order to perform simulations closer to the real soils.
We also intend simulate the rainfall effect over the simulated structure, to 
verify the dependence of the Hurst exponent with the rain \cite{garcia2}. 

This work was sponsored by the brazilian agencies CNPq, FAPEMIG and FINEP, and
by the spanish agency AECI.

\end{document}